\documentstyle[twocolumn,pra,aps,epsfig]{revtex}
\title{Dimensionality effects in restricted bosonic and fermionic systems}
\author{Drago\c s-Victor Anghel}
\address{\em Department of Physics, University of Jyv\"askyl\"a, 
P.O. Box 35, 40351 Jyv\"askyl\"a, Finland,
and NIPNE  -- ``Horia Hulubei'', P.O.BOX MG-6, R.O.-76900 
Bucure\c sti - M\u agurele, Romania}
\begin{document}
\maketitle

\vspace{10mm}
\begin{abstract}
The phenomenon of Bose-like condensation, the continuous change of the 
dimensionality of the particle distribution as a consequence of freezing 
out of one or more degrees of freedom in the limit of low particle density, 
is investigated theoretically in the case of closed systems 
of massive bosons and 
fermions, described by general single-particle hamiltonians. 
This phenomenon is similar for 
both types of particles and, for some energy spectra, exhibits features 
specific to multiple-step 
Bose-Einstein condensation, for instance the appearance of maxima 
in the specific heat. 
In the case of fermions, as the particle density increases, another 
phenomenon is also observed.
For certain types of single particle hamiltonians, the specific heat is 
approaching asymptotically a divergent behavior at zero temperature, as 
the Fermi energy $\epsilon_{\rm F}$ is converging towards any value 
from an infinite discrete set of energies: $\{\epsilon_i\}_{i\ge 1}$. 
If $\epsilon_{\rm F}=\epsilon_i$, for any $i$, the specific heat is 
divergent at $T=0$ just in infinite systems, whereas for any finite system 
the specific heat approaches zero at low enough temperatures.
The results are particularized for particles trapped inside 
parallelepipedic boxes and harmonic potentials.
\end{abstract}

\section{Introduction}

Because of the advances in nanotechnology it has become possible to use 
very small 
structures in a broad range of applications. The importance of these 
applications and the fact that the physical properties of such structures 
could be very different from those of bulk materials, make the theoretical 
and experimental investigations very useful in this area.

The experimental findings in \cite{jukka} motivated us to 
calculate the thermal properties of ultrathin dielectric membranes or wires
by splitting the phonon spectra into discrete and a continuous parts 
\cite{noi1,noi2}. This framework implies 
crossovers between different phonon gas distributions, reflected, for 
example, in the exponent of the temperature dependence of the specific heat or 
heat conductivity. For example in a membrane, as the temperature drops, 
the population of the phonon modes parallel to the  surfaces [which we 
shall call the two-dimensional ground state (2D gs)] becomes 
dominant, and the three-dimensional (3D) phonon gas distribution changes 
into a two-dimensional (2D) one \cite{noi1}. The macroscopic population 
of the 2D gs [or one-dimensional ground state (1D gs) in the 
case of a wire \cite{noi2}] and the qualitative differences between 
phonon gas distributions with various dimensions enabled us to make 
the analogy with the multiple-step Bose-Einstein condensation 
(BEC) \cite{druten,sonin} and to call this phenomenon Bose-like 
condensation (BLC). Yet, the number of phonons changes with temperature 
and features like maxima of the specific 
heat ($c_{\rm V}$) observed in the case of BEC can not be seen in the 
case of a phonon gas undergoing BLC. 

The first purpose of this paper is to extend the previous work 
reported in Refs. \cite{noi1,noi2} and to describe BLC in systems 
of massive bosons and fermions. This will be done in Section \ref{blc}. 
The mathematical technique used 
here is a straightforward extension of the one introduced by Pathria and 
Greenspoon in Ref. \cite{path1}. Nevertheless, the analytical 
approximations used there are not appropriate for our case. 
Therefore, after obtaining general expressions, we make numerical 
calculations to give concrete examples of BLC and to observe 
the behavior of the specific heat during the transition. The phenomenon 
occurs at low particle densities (this will be made more clear in 
Section \ref{blc}) and is specific to both bosons and fermions. 
At low temperatures, the number of massive particles in a closed system 
can be considered to be constant. 
The conservation of the particle number will allow us to observe 
resemblances with the BEC, like, in some cases, maxima of the specific heat 
(${c_{\rm V}}_{\rm max}$) at the condensation temperature. 
Anyway, the signature of BLC, as seen in the temperature 
dependence of $c_{\rm V}$, is more complex and depends on the energy spectrum.

A consequence of the third law of thermodynamics is that the specific 
heat of any thermodynamical system should vanish at zero temperature. 
Li et al. showed in Ref. \cite{li} that the heat capacity of a Fermi gas, 
confined in an external potential of quite general form, and for any 
space dimension, has the asymptotic behavior $c_{\rm V}\propto T$ at 
low temperatures (where $T$ is the temperature of the system). 
This is for the case of a continuous energy spectrum. 
In contrast to this we show in Section \ref{diverg} that 
the specific heat of a Fermi gas with a single-particle hamiltonian of 
the form $H = H_{\rm c} + H_{\rm d}$, with $H_{\rm c}$ having a 
(quasi)continuous spectrum $\epsilon_{\rm c}\in [0,\infty)$ and 
$H_{\rm d}$ having the discrete 
eigenvalues $\epsilon_i$, $i=0,1,\ldots$, may approach, depending on 
the density of the energy levels of $H_{\rm c}$, divergent behavior 
at temperature $T=0$ K as the Fermi energy $\epsilon_{\rm F}$ 
converges to $\epsilon_i$, for any $i\ge 1$. In such a case, if the 
spectrum of $H_{\rm c}$ is continuous, then the specific heat diverges at 
$T=0$ and $\epsilon_{\rm F}=\epsilon_i$, for any $i\ge 1$. However, 
in any finite system the energy spectrum is discrete, so the specific 
heat approaches zero if we go at low enough temperatures and the third 
law of thermodynamics is not violated. 

Ultrathin (semi)conducting membranes and 
wires, nowadays widely used in mesoscopic applications, atoms in very 
anisotropic harmonic traps, wires or constrictions defined in 2D electronic 
gasses are just a few examples of systems where the phenomena presented here
could be observed. Also, they could provide an understanding of the 
behavior of very thin liquid He films. 

\section{Bose-like condensation}\label{blc}

The BEC in cuboidal boxes with small dimensions drew a lot of attention 
many years ago, in the beginning in connection with very thin films 
of liquid He \cite{sonin,path1,krueger,osb-zim,goble1,goble2}. It is 
now well known that, as the dimensions of the box are reduced, at constant 
density, the cusp-like maximum of $c_{\rm V}$ is rounded off 
and the condensation temperature (in this situation taken as the 
temperature corresponding to the maximum) increases with 
respect of the bulk value. The maximum of the specific heat is usually 
smaller in restricted geometries than in the bulk, for all the boundary 
conditions imposed on the walls of the container, with the exception 
(the only one known by the present author) of Dirichlet 
boundary conditions \cite{goble1,goble2,path1}. The theoretical 
investigation of BEC in harmonic traps (see Ref. \cite{holthaus1} 
and references therein) was motivated recently 
by its realization in ultracold trapped atomic gases \cite{a-b-d}. In 
this situation, the specific heat of an infinite system presents a 
discontinuity at the condensation temperature. 
In finite systems, the discontinuity is again rounded off, as shown 
by analytical and numerical calculations, for example in Refs. 
\cite{holthaus,kirsten,haugerud,ketterle}.
As the number of particles is decreased the condensation 
temperature decreases \cite{holthaus,kirsten,haugerud,ketterle}. 
Furthermore, the multiple-step BEC was introduced in Refs. 
\cite{druten,sonin} for the cases of very anisotropic boxes or confining 
potentials. In this case a finite Bose gas is condensing gradually 
to the ground state, exhibiting in between 2D and/or 1D macroscopic 
populations. 

In a very anisotropic Bose system, as particle density decreases, 
the multiple-step BEC (MSBEC) temperature becomes lower than the 
temperature at which some of the degrees of freedom of our system 
freeze out. During both processes (MSBEC and freezing) the 
3D particle distribution transforms gradually into a lower dimensional 
distribution. On the other hand, the two processes change 
into each other at the variation of the particle density or 
of the dimensions of the system. Moreover, 
the reduction of the dimensionality of the particle distribution due to 
the freezing out of some of the degrees of freedom can happen also for 
fermions at low densities. The analogies and differences between the two 
processes mentioned above, justify (arguably, of course) the use 
of the simpler expression of Bose-like condensation for the freesing 
out of degrees of freedom, in the limit of low particle density.

The temperature at which BLC occurs (as in the case of BEC 
in finite systems, this temperature cannot be uniquely 
defined) depends on the energy spectrum and has a finite positive value. 
This type of 
condensation is identical for both bosons and fermions (see Fig. 1). 
To show this, let us consider a closed system of massive bosons and fermions 
described by a single-particle Hamiltonian of the form 
$H = H_{\rm c} + H_{\rm d}$, with the eigenvalues  
$\epsilon=\epsilon_{\rm c}+\epsilon_i$, as explained in the introduction. 
The mean occupation numbers of single particle energy levels $\epsilon$, are 
$\langle n^{(\pm)}_{\epsilon} \rangle = 
(\exp{(\alpha+\epsilon/k_{\rm B}T)}\pm1)^{-1}$, where $(-)$ is the superscript 
for bosons, and $(+)$ for fermions, $\alpha = -\mu/k_{\rm B}T$, and $\mu$ is 
the chemical potential. We introduce the functions 
\begin{eqnarray}
Z_n^{(\pm)}&=&\sum_\epsilon \left(\frac{\epsilon}{k_{\rm B}T}\right)^n\langle 
n_\epsilon^{(\pm)}\rangle ,\label{Zn}\\
G_n^{(\pm)}&=&\sum_\epsilon \left(\frac{\epsilon}{k_{\rm B}T}\right)^n[\langle 
n_\epsilon^{(\pm)}\rangle \mp \langle n_\epsilon^{(\pm)}\rangle^2] 
= -\frac{\partial Z_n^{(\pm)}}{\partial\alpha} ,\label{Gn}
\end{eqnarray}
in a similar way as Pathria and Greenspoon did for bosons in 
\cite{path1}. Then, for example, the number of particles, the 
internal energy, and the heat capacity can be written as 
$N^{(\pm)} = Z_0^{(\pm)}$, $U^{(\pm)} = k_{\rm B}TZ_1^{(\pm)}$, and 
$C_V^{(\pm)} = k_{\rm B}(G_2^{(\pm)}-{G_1^{(\pm)}}^2/G_0^{(\pm)})$, 
respectively (in all this paper we shall consider spinless particles). 
To avoid divergent terms that occur in 
the functions introduced when $T$ approaches zero, in the case when 
the ground state energy $\epsilon_0$ is positive, we redefine $\alpha$ as
$\alpha-\epsilon_0/kT$ and $\epsilon$ as $\epsilon-\epsilon_0$.
Making these replacements we do not change the thermodynamics of 
the canonical ensemble \cite{coment}. If the density of the energy levels 
of the (quasi)continuous spectrum, as a function of energy, is 
$\sigma(\epsilon_{\rm c})$, then we can write 
\begin{equation}
Z_0^{(\pm)}=\sum_{i=0}^\infty\int_0^\infty\frac{\sigma(\epsilon)}{
\exp{(\alpha+\beta\epsilon_i+\beta\epsilon)}\pm 1}\,d\epsilon \, ,\nonumber
\end{equation}
where $\beta=1/k_{\rm B}T$.
If in the temperature range of interest for the study of BLC 
($\epsilon_1/k_{\rm B}T\approx 1$) $\alpha\gg 1$, then we can write $Z_0$ 
in terms of two functions, corresponding to the continuous and to the  
discrete spectra, respectively: 
$$
Z^{(\pm)}_0=e^{-\alpha}Z^{(\pm)}_{\rm c}Z^{(\pm)}_{\rm d} \, ,\nonumber
$$
where $Z^{(\pm)}_{\rm c}=\int_0^\infty\sigma(\epsilon)e^{-\beta\epsilon}\,
d\epsilon$ and $Z^{(\pm)}_{\rm d}=\sum_{i=0}^\infty e^{-\beta\epsilon_i}$. 
Within this approximation is no difference between 
bosons and fermions and, according to Eq. (\ref{Gn}), 
$G^{(\pm)}_n=Z^{(\pm)}_n$. Using the relation
\begin{equation}\label{z0_zn}
\left(\frac{\partial^nZ_n^{(\pm)}}{\partial \alpha^n}\right)_\beta = 
\beta^n\left(\frac{\partial^nZ_0^{(\pm)}}{\partial\beta^n}\right)_\alpha ,
\end{equation}
that holds for bosons \cite{path1}, as well as for fermions, we can write 
the specific heat $c^{(\pm)}_{\rm V}=C^{(\pm)}_{\rm V}/N^{(\pm)}$, in 
units of $k_{\rm B}$, as 
\begin{eqnarray}
\frac{c_{\rm V}}{k_{\rm B}}&=&\beta^2\frac{\partial^2}{\partial\beta^2}
\log{Z_0(\beta,\alpha)} \label{cV}\\
&=&\beta^2\frac{\partial^2}{\partial\beta^2}\log{Z_{\rm c}(\beta)} 
+\beta^2\frac{\partial^2}{\partial\beta^2}\log{Z_{\rm d}(\beta)}\nonumber
\end{eqnarray}
(where we have droped the superscript $(\pm)$ as insignificant in this case). 
Acording to Eq. (\ref{cV}), the specific heat is nothing else then the sum 
of the heat capacities of two systems, each of them containing a single 
particle under canonical conditions, and it is described by the Hamiltonian 
$H_{\rm c}$ and $H_{\rm d}$, respectively.

\begin{figure}[t]
\begin{center}
\unitlength1mm\begin{picture}(80,115)(0,0)
\put(0,0){\epsfig{file=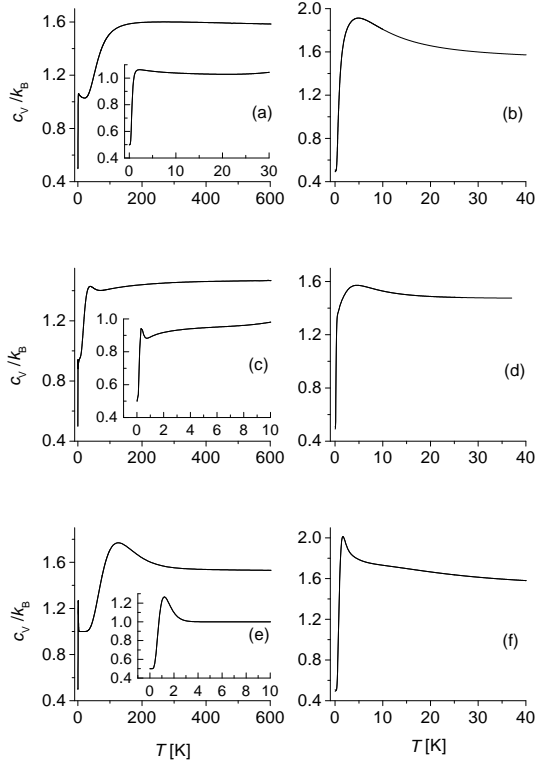,width=80mm}}
\end{picture}
\caption{Specific heat in units of $k_{\rm B}$ vs. temperature for 
ideal Bose and Fermi gases  trapped inside cuboidal boxes  with Dirichlet 
((a) and (b)), Neumann ((c) and (d)), and periodic ((e) and (f)) 
boundary conditions on the walls. 
The Figs. (a), (c), and (e) correspond to $l_1\to\infty$, $l_2=10^{-9}$ m, 
and $l_3=10^{-10}$ m, while the Figs. (b), (d), and (f) correspond to 
$l_1\to\infty$ and $l_2=l_3=10^{-9}$ m. 
The particle density is $10^{25}$ m$^{-3}$ in each case. 
In each situation the results for bosons (solid line) and fermions 
(dashed line) are both plotted, but they can not be distinguished.
In the insets of (a), (c), and (e) we show low temperature details of 
the larger graphs (the axes are the same).}
\end{center}
\end{figure}

Explicit expressions for $Z^{(\pm)}_n$ and $G^{(\pm)}_n$ can be obtained 
if we assume that the density of states of the continuous spectrum has 
the form $\sigma(\epsilon_{\rm c})=C\epsilon_{\rm c}^s$ ($C$ and $s$ are 
constants, such that $C>0$ and $s>-1$), as it happens in most of the 
cases \cite{bagnato}. 
Using the Eqs. (\ref{Zn}), (\ref{Gn}), and (\ref{z0_zn}), we can write:
\begin{eqnarray}
\begin{array}{rcl}
Z_n^{(\pm)}&=&\frac{C}{\beta^{s+1}}\sum_{j=0}^n C_n^j\Gamma(s+1+n-j) \\
& &\times\sum_{i=0}^{\infty} n_i (\beta\epsilon_i)^j g_{s+1+n-j}^{(\pm)}
(\alpha+\beta\epsilon_i) \\ 
{\rm and} & & \\
G_n^{(\pm)}&=&\frac{C}{\beta^{s+1}}\sum_{j=0}^n C_n^j\Gamma(s+1+n-j) \\
& &\times\sum_{i=0}^{\infty} n_i (\beta\epsilon_i)^j g_{s+n-j}^{(\pm)}
(\alpha+\beta\epsilon_i) \, ,
\end{array}\label{zg_n}
\end{eqnarray}
where $n_i$ is the degeneracy of the level with energy $\epsilon_i$ and 
$C_k^j=n!/j!(n-j)!$. The functions $g^{(\pm)}_l(\alpha)$ are the 
$l^{\rm th}$ order polylogarithmic functions (see for example Ref. 
\cite{lee} and the references therein for more details) of argument 
$e^{-\alpha}$ (bosons) or $-e^{-\alpha}$ (fermions). In the case of ideal 
particles inside a rectangular box of dimensions 
$l_{\rm x}\gg l_{\rm y},l_{\rm z}$, we can write 
$\epsilon_{\rm c}=\hbar^2k_{\rm x}^2/2m$ and 
$\epsilon_{\{i,j\}}=\hbar^2(k_{{\rm y}i}^2+k_{{\rm z}j}^2)/2m$, where 
$k_{\rm x}$, $k_{\rm y}$, and $k_{\rm z}$ are the wave vectors along the 
$x$, $y$, and $z$ axes, respectively. The mass of one particle is $m$ and 
the discrete values of $k_{{\rm y}i}$ and $k_{{\rm z}j}$ depend on the 
boundary conditions. In this case $s=-1/2$. If 
$l_{\rm x},l_{\rm y}\gg l_{\rm z}$, then $s=0$ and 
$\epsilon_{\rm c}=\hbar^2(k_{\rm x}^2+k_{\rm y}^2)/2m$, while 
$\epsilon_{i}=\hbar^2k_{{\rm z}i}^2/2m$. 
Let us now concentrate on the BLC of particles inside such rectangular boxes. 
In Fig. 1 we can see the results of the exact numerical calculation 
of $c_{\rm V}$ (using the formulae from Eq. (\ref{zg_n}) for 
$Z^{(\pm)}_n$ and $G^{(\pm)}_n$) as a function of temperature, for two 
different kinds of geometries and for Dirichlet (Fig. 1 (a), (b)), 
Neumann (Fig. 1 (c), (d)), and periodic (Fig. 1 (e), (f)) boundary 
conditions. In geometry I (see Fig. 1 (a), (c), and (e)) $l_2=10^{-9}$ m, 
$l_3=10^{-10}$ m, and $l_1\gg l_2$, while in geometry II (see Fig. 1 (b), 
(d), and (f)) $l_2=l_3=10^{-9}$ m and $l_1\gg l_2$. To make concrete 
calculations we choose 
$\lambda^2\equiv 2\pi\hbar^2/mk_{\rm B}T=10^{-18}T^{-1}$ which corresponds to 
a mass of about 3 atomic mass units for all the particles in the systems 
investigated. In the figure, the results for bosons and fermions are 
indistinguishable, as expected for low particle densities. The choice of 
the dimensions in geometry I allow us to observe the BLC from 3D to 2D and, 
at lower temperature, from 2D to 1D. We observe the formation of a maximum 
(at, let us say, temperature $T_{\rm max}$) in each of these two cases and 
for all boundary conditions. The height of this maximum and, 
in general, the shape of the function $c_{\rm V}(T)$ around 
$T_{\rm max}$ depend on the spectrum of $H_{\rm d}$. For example, for 
Neumann boundary conditions, we observe the formation of a minimum at 
a temperature a bit higher than $T_{\rm max}$. In geometry II we 
observe the BLC from 3D to 1D. In this case, the maxima are more pronounced 
and the minima observed in geometry I for Neumann boundary conditions 
disappear. 

In Fig. 2 we plot $T_{\rm max}/T_{\rm c}$ and 
${c_{\rm V}}_{\rm max}/k_{\rm B}$ 
vs. $l_3/l$, for Dirichlet, Neumann, and periodic boundary conditions, in 
the cases when $l_1,l_2\gg l_3$ (Fig. 2 (a), (c)) and $l_1\gg l_2=l_3$ 
(Fig. 2 (b), (d)). $T_{\rm c}$ is the bulk BEC temperature, given by the 
equation $\rho(2\pi\hbar^2/mk_{\rm B}T_{\rm c})^{3/2}=\zeta(3/2)$, $\rho$ 
is the particle density, $\zeta$ is the Riemann zeta function, and 
$l=\rho^{-1/3}$ is the mean interparticle distance. 
Since $T_{\rm max}$ converges to a finite value and $T_{\rm c}\to 0$ 
when $\rho \to 0$, the ratio $T_{\rm max}/T_{\rm c}$ diverges in this limit. 
As $\rho$ increases, $T_{\rm c}$ increases and BLC is gradually replaced by 
BEC. As a consequence, $\lim_{l_3/l\to\infty}T_{\rm max}/T_{\rm c}=1$.
Fig. 2 (c) can make the connection between these numerical calculations 
and the analytical approximations reported in Ref. \cite{path1}. 
We observe that ${c_{\rm V}}_{\rm max}$ is higher in 
Fig. 2 (d) then in Fig. 2 (c), at the same value of $l_3/l$, 
for any boundary conditions. Nevertheless, the maximum value of 
${c_{\rm V}}_{\rm max}$, which is about $2.02k_{\rm B}$,  
is obtained for periodic boundary conditions in the limit $\rho\to 0$, 
while at higher densities this decreases under its bulk value, as 
expected from previous calculations  \cite{path1}.

\begin{figure}[t]
\begin{center}
\unitlength1mm\begin{picture}(80,70)(0,0)
\put(0,0){\epsfig{file=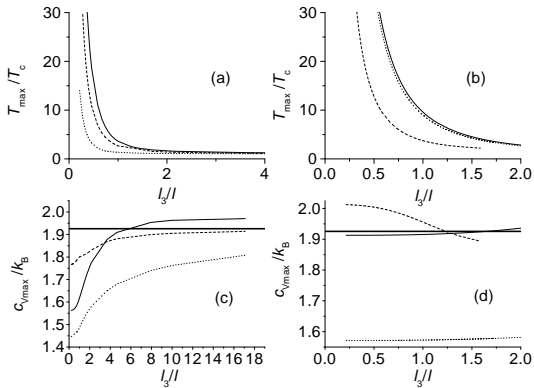,width=80mm}}
\end{picture}
\caption{The temperature of ${c_{\rm V}}_{\rm max}$ 
of {\em Bose gases}, scaled by the bulk critical 
temperature $T_c$ (see the text), as a function of $l_3/l$ 
(where $l=\rho^{-1/3}$, and 
$\rho$ is the density), is shown for (a) the membrane geometry 
($l_1, l_2\gg l_3$) and (b) the wire geometry ($l_1\gg l_2=l_3$).
The value of ${c_{\rm V}}_{\rm max}$, in units of $k_{\rm B}$, vs. $l_3/l$, 
is plotted for (c) membrane geometry and (d) wire geometry (the same as in 
(a) and (b)). Solid, dashed and dotted lines are used for Dirichlet, 
periodic and Neumann boundary conditions. The thick horizontal lines in 
(c) and (d) correspond to the 3D bulk value of $c_{\rm V}$ at the BEC 
temperature. In the numerical calculations we varied $\rho$, keeping 
$l_3=10^{-9}$ m. }
\end{center}
\end{figure}

The study the BLC of ideal particles in harmonic traps is easier since 
in this situation $Z_{\rm d}$ has a very simple analytical expression. 
If we denote 
the characteristic frequencies of the harmonic trap by $\omega_{\rm x}$, 
$\omega_{\rm y}$, and $\omega_{\rm z}$, with 
$\omega_{\rm x}\ll\omega_{\rm y},\omega_{\rm z}$, then 
$Z_{\rm c}=k_{\rm B}T/\hbar\omega_{\rm x}$ and 
$Z_{\rm d}=[(1-\exp{(\hbar\omega_{\rm x}/k_{\rm B}T)})
(1-\exp{(\hbar\omega_{\rm y}/k_{\rm B}T)})]^{-1}$.
In this case $dc_{\rm V}/dT\ge 0$ for any temperature, so BLC is not 
accompanied by the formation of a maximum. The dimensionality 
of the system (say, $n$D) is reflected in the value of $c_{\rm V}$, which is 
$nk_{\rm B}$, and the fraction of the
particle number in the 1D gs, has the expression $N_{\rm 1D}/N=
(1-e^{-(\hbar\omega_{\rm y}/k_{\rm B}T)})
(1-e^{-(\hbar\omega_{\rm z}/k_{\rm B}T)})$.

\section{Divergent behavior of $c_{\rm V}$ in fermionic systems}
\label{diverg}

In this section we shall concentrate on Fermi systems close to $T=0$ K. 
We consider again that the Hamiltonian of the system can be approximated 
by single-particle operators of the form $H = H_{\rm c} + H_{\rm d}$, as 
explained in the introduction. At the increase of the particle density 
or of the density of the eigenvalues $\epsilon_i$ of the operator 
$H_{\rm d}$, we would expect to approach the limit in which both, 
$H_{\rm c}$ and $H_{\rm d}$ have continuous spectra (3D bulk limit). 
In such a limit we should recover the results from 
Ref. \cite{li}, namely $c_{\rm V}\propto T$ at low temperatures. 
As it will be shown next, 
this is not the case in general. The continuous limit is not attained 
in a smooth way. Instead, in some situations, the specific heat would become 
divergent at zero temperature, for certain values of the Fermi energy.

At temperatures close to 0 K the 
chemical potential of a Fermi system approaches the Fermi energy 
$\epsilon_{\rm F}$. For 
$\alpha\ll -1$, the polylogarithmic functions of negative argument 
can be written in the form \cite{li}:
\begin{equation}\label{rl}
g^{(+)}_n(\alpha)=\frac{|\alpha|^n}{\Gamma(n+1)}\left[1+{\mathcal O}\left(
\frac{1}{\alpha^2}\right)\right] \, . 
\end{equation}
The cases for $n=$0 and 1 are included in (\ref{rl}), but can be 
refined further 
to write $g^{(+)}_n(\alpha)=|\alpha|^n\left[1+{\mathcal O}
\left(e^{\alpha}\right)\right]$. In the other extreme case, when 
$\alpha\gg 1$, all the polylogarithmic functions have a behavior of the 
form  $g^{(\pm)}_n(\alpha)=e^{-\alpha}\left[1+{\mathcal O}
\left(e^{-\alpha}\right)\right]$. Using these asymptotic expressions we can 
return to the study of the specific heat close to zero temperature, 
for a density of energy levels of $H_{\rm c}$ similar to the one 
introduced in the previous section, namely 
$\sigma(\epsilon_{\rm c})=C\epsilon_{\rm c}^s$. The ground state of 
$H_{\rm d}$ is nondegenerate since we discuss a finite 
system. We shall use the notation 
$\alpha_0\equiv -\beta\epsilon_{\rm F}$. 

Since we know that $\mu\to\epsilon_{\rm F}$ as $T\to 0$, let us now calculate 
$\lim_{T\to 0}(|\alpha_0|-|\alpha|)$ when $\epsilon_{\rm F}=\epsilon_i$, 
$i>0$ (in all the other cases will turn out that the limit is zero).
Using $N=Z^{(+)}_0(\alpha)$, Eqs. (\ref{zg_n}), and the definition of the 
Fermi energy, we write two different expressions for the total number of 
particles in the system:
\begin{eqnarray}
N&=&\frac{C}{(s+1)\beta^{s+1}}\left\{|\alpha|^{s+1}+\ldots +
(|\alpha|-\beta\epsilon_{i-1})^{s+1}\right\}\nonumber \\
& &\times\left[1+{\mathcal O}\left(
\frac{1}{\alpha^2}\right)\right] \label{n1} \\
& &+n_iC\frac{\Gamma(s+1)}{\beta^{s+1}}g^{(+)}_{s+1}(\alpha+\beta\epsilon_i)
\nonumber \\
& &+ C\frac{\Gamma(s+1)}{\beta^{s+1}}\sum_{j=i+1}^{\infty}n_j 
e^{|\alpha|-\beta\epsilon_j} \nonumber \\
 &=& \frac{C}{(s+1)\beta^{s+1}}\left\{|\alpha_0|^{s+1}+\ldots +
(|\alpha_0|-\beta\epsilon_{i-1})^{s+1}\right\} \, . \label{n2}
\end{eqnarray}
If we denote $\xi\equiv\alpha+\beta\epsilon_i$, then from (\ref{n1}) and 
(\ref{n2}), neglecting the exponentials and assuming that 
$\lim_{T\to 0}(\xi/|\alpha_0|)=\lim_{T\to 0}(\xi/|\alpha|)=0$, 
we obtain, in the case $\alpha_0,\alpha\ll -1$, an equation for $\xi$:
\begin{equation}\label{xi}
n_i\frac{g^{(+)}_{s+1}(\xi)}{\xi}=\frac{|\alpha_0|^s}{\Gamma(s+1)}\chi_s\, , 
\end{equation}
where $\chi_s\equiv 1+\ldots+n_{i-1}(1-x_{i-1})^s$ and 
$x_j\equiv\epsilon_j/\epsilon_{\rm F}$. 
We now notice that we have three distinct situations: ({\em a}) $s>0$, in 
which case $\xi\to 0$ as $T\to 0$, ({\em b}) s=0, and $\xi$ converges to a 
finite positive value, and ({\em c}) $s\in(-1,0)$, when 
$\xi\to\infty$ as $T\to 0$. 

Let us now analyze the asymptotic behavior of $\xi$ in the case, 
({\em c}). For $\xi \gg 1$ we can write 
\begin{equation}\label{xi_asimpt}
\frac{e^{-\xi}}{\xi}=\frac{|\alpha_0|^s}{n_i\Gamma(s+1)}\chi_s\, , 
\end{equation}
so $\xi=(-s)\log{|\alpha_0|}-\log{\xi}-\log{(\chi_s/n_i\Gamma(s+1))}$. 
Therefore, at $\alpha_0\ll -1$, $\xi \approx |s|\log{|\alpha_0|}-
\log{[\log{|\alpha_0|}]}+\ldots$. We can see now that the assumption 
$\lim_{T\to 0}(\xi/|\alpha_0|)=\lim_{T\to 0}(\xi/|\alpha|)=0$ was 
justified. Also, following the same kind 
of reasoning, one can prove that when $\epsilon_{\rm F}\neq\epsilon_i$, 
for any $i$, then $\lim_{T\to 0}(|\alpha_0|-|\alpha|)=0$ for any $s$.

Using the Eqs. (\ref{xi}) and (\ref{xi_asimpt}) we can 
calculate the specific heat close to 0 K. For that we have to
evaluate the functions $G^{(+)}_2$, $G^{(+)}_1$, $G^{(+)}_0$, and 
$Z^{(+)}_0$. We analyze again the case when $\epsilon_{\rm F}=\epsilon_i$, 
$i>0$. After some algebra and dropping out the factors that become 
exponentially small in the limit $T\to 0$, we can write: 
\begin{eqnarray}
G^{(+)}_2&=&\frac{C|\alpha|^{s+2}}{\beta^{s+1}}\left\{\chi_s
\left[1+{\mathcal O}\left(\frac{1}{\alpha^2}\right)\right]\right.
\label{g2} \\
 & & \left.+n_i 
\left[\Gamma(s+3)\frac{g^{(+)}_{s+2}(\xi)}{|\alpha|^{s+2}}+2\Gamma(s+2)y_i 
\frac{g^{(+)}_{s+1}(\xi)}{|\alpha|^{s+1}}\right.\right. \nonumber \\
& &\left.\left. +\Gamma(s+1)y^2_i 
\frac{g^{(+)}_{s}(\xi)}{|\alpha|^{s}}\right] - 
\frac{s\xi}{|\alpha|}\Upsilon_s\right\} , \nonumber \\
G^{(+)}_1&=&\frac{C|\alpha|^{s+1}}{\beta^{s+1}}\left\{\chi_s
\left[1+{\mathcal O}\left(\frac{1}{\alpha^2}\right)\right]\right. \label{g1}\\
& &\left.+n_i 
\left[\Gamma(s+2)\frac{g^{(+)}_{s+1}(\xi)}{|\alpha|^{s+1}}+\Gamma(s+1)y_i 
\frac{g^{(+)}_{s}(\xi)}{|\alpha|^{s}}\right] \right. \nonumber\\
& &\left. - \frac{s\xi}{|\alpha|}\Upsilon_s\right\} , \nonumber\\
G^{(+)}_0&=&\frac{C|\alpha|^{s}}{\beta^{s+1}}\left\{\chi_s
\left[1+{\mathcal O}\left(\frac{1}{\alpha^2}\right)\right]\right.
\label{g0} \\
& &\left.+n_i\Gamma(s+1)\frac{g^{(+)}_{s}(\xi)}{|\alpha|^{s}} - 
\frac{s\xi}{|\alpha|}\Upsilon_s\right\} , 
\nonumber\\
Z^{(+)}_0&=&\frac{C|\alpha|^{s+1}}{(n+1)\beta^{s+1}}\left\{\chi_{s+1}
\left[1+{\mathcal O}\left(\frac{1}{\alpha^2}\right)\right]\right.\label{z0}\\
& &\left.+n_i 
\Gamma(s+2)\frac{g^{(+)}_{s+1}(\xi)}{|\alpha|^{s+1}} - 
\frac{(s+1)\xi}{|\alpha|}\Upsilon_{s+1}\right\} , \nonumber
\end{eqnarray}
where $y_j=\beta\epsilon_j/|\alpha|$ and $\Upsilon_s \equiv 
\sum_{k=1}^{i-1} n_k(1-x_k)^{s-1}x_k$. 
To see the asymptotic behavior, we calculate $c_{\rm V}$ separately 
for the cases ({\em a}), ({\em b}), and ({\em c}). 
Using Eqs. (\ref{xi},\ref{g2}-\ref{z0}) and working consistently in 
the orders of $|\alpha|$, we obtain the following asymptotic results:
\begin{itemize}
\item {\em Case (a)}
\begin{eqnarray}
\frac{c_{\rm V}}{k_{\rm B}}&=&\frac{(s+1)|\alpha|}{\chi_{s+1}+
{\mathcal O}(|\alpha|^{-(s+1)})} \label{cv_a} \\
 & &\times\left\{\frac{n_i^2\Gamma^2(s+1)}{\chi_s}
\frac{{g_s^{(+)}}^2(\xi)}{|\alpha|^{2s}}\right.\nonumber\\
& &\left.+
\frac{n_i^3\Gamma^3(s+1)}{\chi_s^2}
\frac{{g_s^{(+)}}^3(\xi)}{|\alpha|^{3s}} + {\mathcal O}\left(
\frac{1}{|\alpha|^m}\right)\right\}\, , \nonumber
\end{eqnarray}
where $m = \min{\{s+1,4s,2\}}$.
\item {\em Case (b)}
\begin{eqnarray}
\frac{c_{\rm V}}{k_{\rm B}}&=&\frac{n_i}{|\alpha|\left(\chi_{1}+
|{\mathcal O}(|\alpha|^{-1})\right)} \label{cv_b} \\
 & &\times\left\{\frac{\chi_0g_0^{(+)}(\xi)}{\chi_0+n_ig_0^{(+)}(\xi)}\xi^2+
\frac{2\chi_0g_1^{(+)}(\xi)}{\chi_0+n_ig_0^{(+)}(\xi)}\xi\right.\nonumber\\
& &\left.+2g_2^{(+)}(\xi)-
\frac{n_i{g_1^{(+)}}^2(\xi)}{\chi_0+n_ig_0^{(+)}(\xi)} \right. \nonumber \\ 
& &\left. + \frac{\pi^2}{3}\chi_0 + {\mathcal O}
\left(e^\alpha\right)\right\}\, ,\nonumber
\end{eqnarray}
\item {\em Case (c)}
\begin{eqnarray}
\frac{c_{\rm V}}{k_{\rm B}}&=&\frac{s+1}{\chi_{s+1}+
{\mathcal O}(|\alpha|^{-(s+1)})}\frac{|\alpha|}{\xi} 
\nonumber\\
& &\times\left\{\chi_s+\frac{\chi_s}{\xi}+{\mathcal O}
\left(\frac{1}{|\alpha|}\right)\right\}\, .\label{cv_c}
\end{eqnarray}
\end{itemize}
So, for $\epsilon_{\rm F}=\epsilon_i$, $i>0$, from Eqs. 
(\ref{cv_a}-\ref{cv_c}) we distinguish the following situations:
\begin{description}
\item[({\em a1})] $s>1/2$, then 
$c_{\rm V}/k_{\rm B}\propto (\epsilon_{\rm F}/k_{\rm B}T)^{1-2s}$, so 
$c_{\rm V}\to 0$ as $T\to 0$ (note that 
if $s>1$ some of the orders of $\alpha$ interchange, but the function 
$c_{\rm V}$ converges fast to zero as $T$ approaches 0 K);
\item[({\em a2})] $s=1/2$, then $\lim_{T\to 0}(c_{\rm V}/k_{\rm B})=
(3-2\sqrt{2})(3\pi/8)\zeta^2(1/2)n_i^2/\chi_{3/2}\chi_{1/2}$; 
\item[({\em a3})] $s\in (0,1/2)$, then 
$c_{\rm V}/k_{\rm B}\propto (\epsilon_{\rm F}/k_{\rm B}T)^{1-2s}$, so 
$c_{\rm V}\to\infty$ as $T\to 0$;
\item[({\em b})] $s=0$, then 
$c_{\rm V}/k_{\rm B}\propto k_{\rm B}T/\epsilon_{\rm F}$, so 
$c_{\rm V}\to 0$ as $T\to 0$;
\item[({\em c})] $s\in (-1,0)$, then 
$c_{\rm V}/k_{\rm B}\propto 
(\epsilon_{\rm F}/k_{\rm B}T)/\log{(\epsilon_{\rm F}/k_{\rm B}T)}$, so 
$c_{\rm V}\to\infty$ as $T\to 0$.
\end{description}
Therefore, in the cases ({\em a3}) and ({\em c}), $c_{\rm V}$ presents a 
divergent behavior at $T=0$ K, while in case ({\em a2}) approaches a finite 
limit. These situations seem to be in contradiction with the third law 
of thermodynamics. To clarify this we mention that the divergency appears
just if the spectrum of $H_{\rm c}$ is continuous. In any finite system 
this is not the case, so at low enough temperatures $c_{\rm V}$ decreases 
towards zero. 

Without getting into details we state that when 
$\epsilon_{\rm F}\neq\epsilon_i,\ \forall i\ge 0$, similar calculations 
lead us to the results $\lim_{T\to 0}(|\alpha_0|-|\alpha|) = 0$ and 
$\lim_{T\to 0}c_{\rm V}=0$ for any $s$. Moreover, in the low 
temperature limit we reobtain the known result \cite{li} 
$c_{\rm V}\propto T$. On the other hand, the 
continuity of $\mu$ as a function of $\epsilon_{\rm F}$ implies the 
continuity of $\alpha$ and $c_{\rm V}$ as functions of $\epsilon_{\rm F}$, 
for any $T>0$ K. In other words, the divergent behavior in the cases 
({\em a3}) and ({\em c}) can be approached asymptotically for any $T>0$ K, 
as $\epsilon_{\rm F}\to\epsilon_i$ (for any $i$), by the 
functions $c_{\rm V}(T)$. This leads to the formation of a maximum at 
finite temperature, with the properties: ${c_{\rm V}}_{\rm max}\to\infty$ and 
$T_{\rm max}\to 0$, as $\epsilon_{\rm F}\to\epsilon_i$, for any $i\ge 1$. 

\begin{figure}[t]
\begin{center}
\unitlength1mm\begin{picture}(80,60)(0,0)
\put(0,0){\epsfig{file=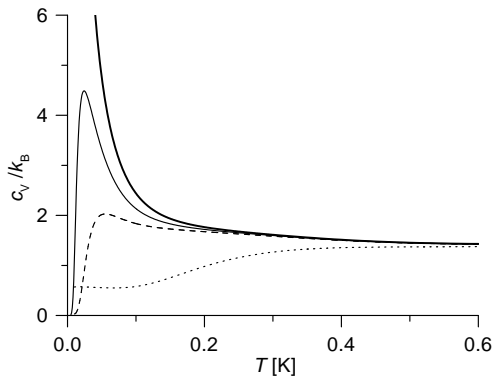,width=80mm}}
\end{picture}
\caption{The specific heat (in units of $k_{\rm B}$) of a Fermi gas 
trapped inside a cuboidal box ($l_1\to\infty,\ l_2=l_3=10^{-9}$ m) 
with Neumann boundary conditions on the walls. The four curves 
correspond to the following densities: $1.5\times 10^{26}$ m$^{-3}$ 
(dotted line), $9.2\times 10^{26}$ m$^{-3}$ (dashed line), 
$9.6\times 10^{26}$ m$^{-3}$ (solid line), and 
$1\times 10^{27}$ m$^{-3}$ (thick solid line). This last case correspond to 
$\epsilon_{\rm F}=\epsilon_1$.}
\end{center}
\end{figure}

Let us now make the connections with familiar systems, namely with 
the ones discussed in section \ref{blc}. In the case of a cuboidal box 
with dimensions $l_{\rm x}\gg l_{\rm y},l_{\rm z}$, $s=-1/2$, so we are in 
the case ({\em c}). In Fig. 3 we plot the exact numerical calculation 
of such a fermionic system, with dimensions 
$l_1\to\infty,\ l_2=l_3=10^{-9}$ m. The mass of the particles is chosen 
as in Section \ref{blc}, such that $\lambda^2=10^{-18}T^{-1}$. We observe the
formation of the maximum as the Fermi energy approaches the first excited 
energy level of $H_{\rm d}$, and the divergent behavior at 
$\epsilon_{\rm F}=\epsilon_1$. If the fermions are inside a cuboidal box 
with dimensions $l_{\rm x},l_{\rm y}\gg l_{\rm z}$ or a harmonic 
potential with the characteristic frequencies 
$\omega_{\rm x}\ll\omega_{\rm y},\omega_{\rm z}$, then $s=0$ and we are in the 
case ({\em b}), therefore we do not observe the formation of a similar 
maximum. This was checked by exact numerical calculations and was found 
to be correct.

\section{Conclusions}

In Section \ref{blc} of this paper it is presented in general 
the phenomenon of {\em Bose-like condensation} in the case of 
massive bosons and fermions. 
This denomination was introduced in \cite{noi1} where, according to my 
knowledge, it was reported for the first time a crossover between 
different dimensionalities of the phonon gas distribution in 
ultrathin dielectric membranes. This phenomenon appears to be identical 
for both types of massive particles and resembles to 
the multiple-step Bose-Einstein condensation \cite{sonin,druten}. 
Nevertheless, the two phenomena are different in nature. 
The results are exemplified for the familiar cases of ideal particles 
trapped inside cuboidal boxes and harmonic potentials.

The analysis made in Section \ref{diverg}, lead us to the 
observation of interesting divergences of the specific heat of a 
Fermi system at zero temperature. The phenomenon 
is described in general, for a single-particle hamiltonian of the form 
$H = H_{\rm c} + H_{\rm d}$, with $H_{\rm c}$ having a 
(quasi)continuous spectrum $\epsilon_{\rm c}\in [0,\infty)$ with the 
energy levels density $\sigma(\epsilon_{\rm c})=C\epsilon_{\rm c}^s$ 
($s>-1$) and 
$H_{\rm d}$ having the discrete eigenvalues $\epsilon_i$, $i=0,1,\ldots$. 
It was found that $c_{\rm v}(T)\to\infty$ as $T\to 0$ for any 
$s\in (-1,0)\cup (0,1/2)$ 
if $\epsilon_{\rm F}=\epsilon_i$, for any $i\ge 1$. This divergent 
behavior is approached asymptotically for any $T>0$, as 
$\epsilon_{\rm F}\to\epsilon_i,\ \forall i\ge 1$, leading in this way to the 
formation of very high maxima (in the limit, infinitely high) of the 
fermionic specific heat close to zero temperature.
This is an unexpected new phenomenon, since it seems to contradict the 
third law of thermodynamics. Anyway, this does not happen since 
in any finite system the energy spectrum is discrete and at low enough 
temperature the specific heat decreases towards zero. Nevertheless, this 
phenomenon might have interesting consequences on the entropy of the 
system in the vicinity of zero temperature. On the other hand it 
should be investigated if systems obeying fractional-statistics \cite{haldane} 
or interacting Bose systems (see for example Ref. \cite{bhaduri,lee1} and 
references therein for similarities between these two types of 
systems) exhibit similar behavior. 

The author wants to thank Professors M. Manninen, J. P. Pekola, and  
E. B. Sonin for discussions. 
This work has been supported by the Academy of Finland under the Finnish
Centre of Excellence Programme 2000-2005 (Project No. 44875, Nuclear and
Condensed Matter Programme at JYFL).




\begin{thebibliography}{15}
\bibitem{coment} The bosonic systems with non-zero ground state energy 
create one more problem to the grand-canonical ensemble, besides the 
well known one related to the fluctuations of the number of particles 
of a Bose-Einstein condensed system \cite{ziff,holthaus1}.
To achieve BEC in such a system ($\alpha \to 
-\epsilon_0/kT$) one should request the ground state energy of the 
reservoir to be greater than or equal to the ground state energy of 
the system under investigation. This implies restrictions on the 
geometry or on the trapping potential of the reservoir.
\end{thebibliography}
\end{document}